\documentclass[aps,pra,groupedaddress,showpacs,showkeys,twocolumn]{revtex4}
\usepackage{natbib}
\usepackage[latin1]{inputenc}
\usepackage[dvips]{graphicx}
\usepackage{amsmath,amsbsy,amsfonts,amssymb}
\usepackage{bm}

\newcommand{\ket}[1]{\ensuremath{\left|#1\right>}}
\newcommand{\cafp}{\ensuremath{^{40}Ca^{+}}}
\newcommand{\omz}{\ensuremath{\omega_{z}}}
\newcommand{\omr}{\ensuremath{\omega_{r}}}
\newcommand{\state}[3]{\ensuremath{\,^{#1}{#2}_{#3}}}
\newcommand{\unit}[1]{\ensuremath{\,\rm #1}}
\begin{document}
\title{Lifetime measurement of the metastable $3d\state{2}{D}{5/2}$ state in the \cafp\ ion using the shelving technique on a few-ion string }
\author{Peter Staanum~\footnote{E-mail: staanum@phys.au.dk}}
\author{Inger S. Jensen}
\author{Randi G. Martinussen}
\author{Dirk Voigt~\footnote{Now at Huygens Laboratory, University of Leiden, The Netherlands.}}
\author{Michael Drewsen}
\affiliation{QUANTOP - Danish National Research Foundation Center
for Quantum Optics, Department of Physics and Astronomy,
University of Aarhus, DK-8000 Aarhus C, Denmark.}
\date{\today}
\begin{abstract}
We present a measurement of the lifetime of the metastable
$3d\state{2}{D}{5/2}$ state in the \cafp\ ion, using the so-called
shelving technique on a string of five Doppler laser-cooled ions
in a linear Paul trap. A detailed account of the data analysis is
given, and systematic effects due to unwanted excitation processes
and collisions with background gas atoms are discussed and
estimated. From a total of 6805 shelving events, we obtain a
lifetime $\tau=1149\pm 14(\rm stat.)\pm 4(\rm sys.)\unit{ms}$, a
result which is in agreement with the most recent measurements.

\end{abstract}
 \maketitle
\section{Introduction}
The metastable $3d\state{2}{D}{5/2}$ state in \cafp\ is
interesting for such diverse areas of physics as atomic structure
calculations, optical frequency standards, quantum information and
astronomy. For atomic structure calculations it is an important
test case for the study of valence-core interactions and
core-polarization effects~\citep{Biemont-Zeippen}. The long
lifetime of the \state{2}{D}{5/2} state implies a sub-Hz natural
linewidth of the 729\unit{nm} electric quadrupole transition to
the \state{2}{S}{1/2} ground state, thus making this transition an
attractive candidate for an optical frequency
standard~\citep{Madej-freq.standard, Knoop-lifetimeII}. In
addition, its long lifetime makes the \state{2}{D}{5/2} state a
suitable choice for one of the two states in a quantum bit in
connection with quantum computation~\citep{Schmidt-Kaler-JPhysB}.
Finally, in astronomy, the \state{2}{D}{5/2} state has been used
in a study of the $\beta$ pictoris disk~\citep{Hobbs-CaInStars},
and it is also commonly used in the study of Seyfert 1 galaxies
and T Tauri stars~\citep{Zeippen-lifetime}. As a consequence, the
natural lifetime of the \state{2}{D}{5/2} state has attracted much
attention in recent years, but unfortunately the results of the
many measurements~\citep{Barton, Block, Lidberg-lifetime,
Ritter-lifetime, Gudjons-lifetime, Knoop-lifetime,
Arbes-lifetimeII, Donald-correlations, Knoop-lifetimeII} and
theoretical calculations~\citep{Biemont-Zeippen, Liaw-lifetime,
Brage-lifetime, Vaeck-lifetime, Guet-lifetime, Zeippen-lifetime,
Ali-Kim} of the lifetime are scattered over a rather broad range
(see Fig.~1 in Ref.~\citep{Barton} for an overview). Among the
experimental results, the lifetime found by Barton \textit{et
al.}~\citep{Barton}, using the so-called shelving technique on a
single trapped and laser-cooled \cafp\ ion, has the smallest error
bars, with an estimated lifetime of $\tau=1168\pm 7\unit{ms}$.

In this paper we present a measurement of the lifetime of the
$3d\state{2}{D}{5/2}$ state, using the same shelving technique but
on a string of five ions. The measurement results in a lifetime of
$\tau=1149\pm14(\rm stat.)\pm4(\rm sys.)\unit{ms}$.

The experimental setup and the experimental procedure are
described in Section~\ref{sec:Experimental}. In
Section~\ref{sec:Data Analysis}, we give a detailed account of the
data analysis, including a description of the maximum likelihood
method used for the statistical data analysis and a discussion of
systematic errors with emphasis on radiation effects and collision
effects. Finally, in Section~\ref{sec:conclusion}, the estimated
lifetime based on the measurements is given and discussed.
\section{Experimental setup}\label{sec:Experimental}
\begin{figure}
  \centering
  \includegraphics[width=\linewidth]{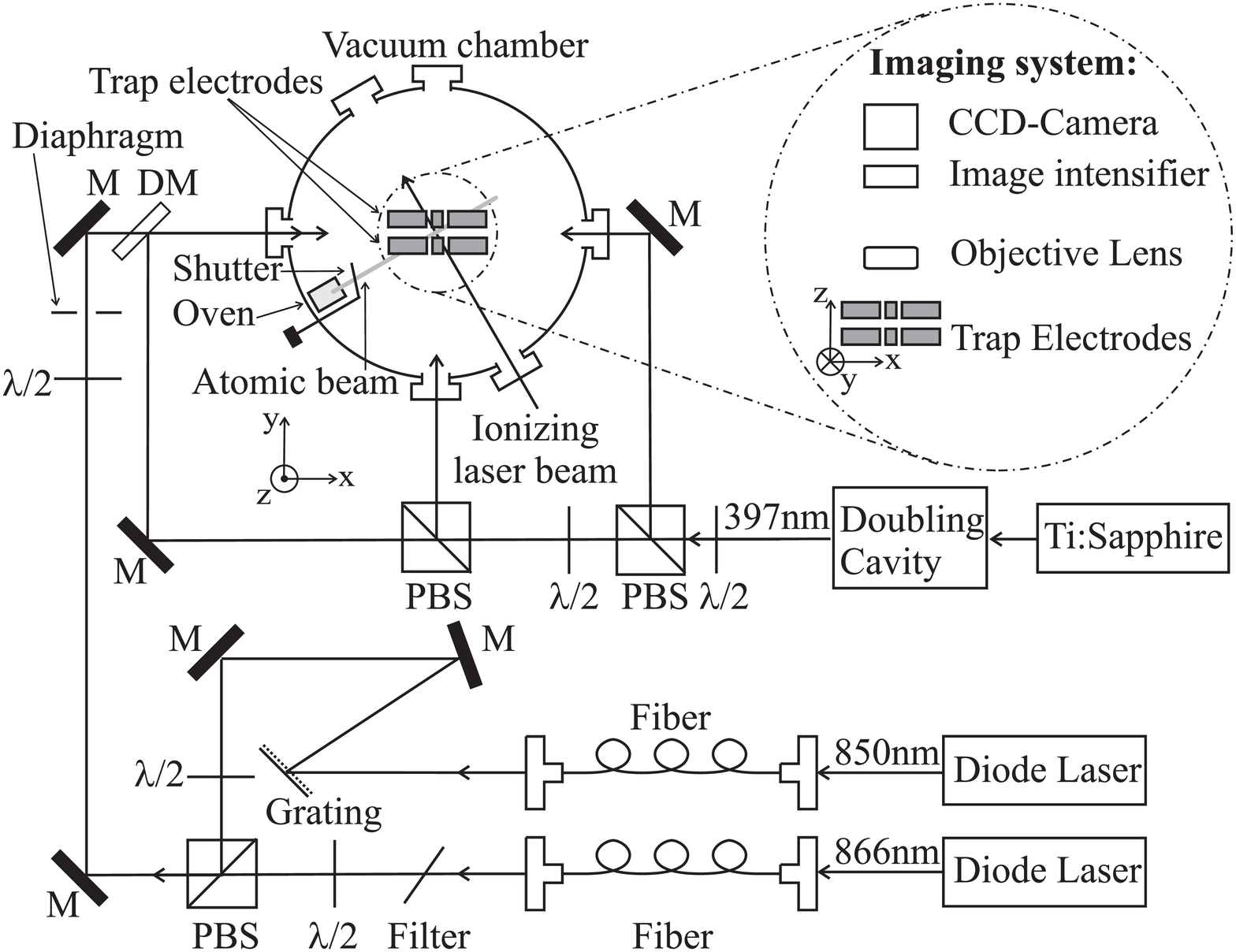}
  \caption{Experimental setup, see text for details. M: mirror, DM: dichroic mirror,
   PBS: polarizing beamsplitter, $\lambda/2$: half-wave plate.}\label{fig:setup}
\end{figure}

\begin{figure}
  \centering
  \includegraphics[width=\linewidth]{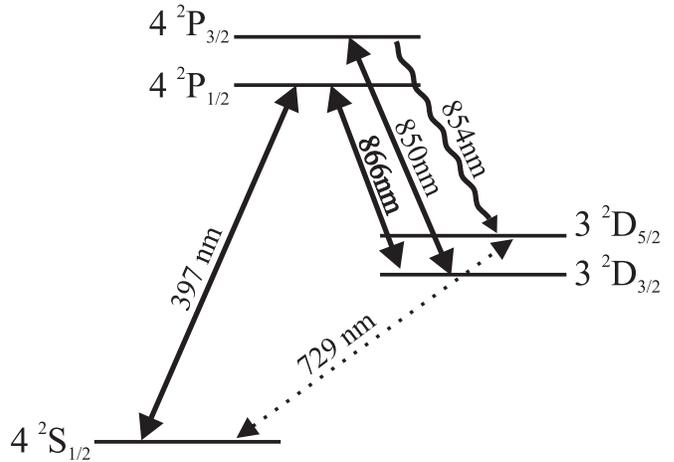}
  \caption{Relevant levels and transitions in \cafp. Doppler laser cooling is performed with lasers at 397\unit{nm} and 866\unit{nm}. A laser at 850\unit{nm}
  excites the ion from the $3d\state{2}{D}{3/2}$ state to the $4p\state{2}{P}{3/2}$ state, from where it can decay to the $3d\state{2}{D}{5/2}$ state.
  The dashed line indicates the electric quadrupole transition by which the ion can decay from the \state{2}{D}{5/2} state to the ground state.}\label{fig:optical-pumping}
\end{figure}

\begin{figure}
  \centering
  \includegraphics[width=\linewidth]{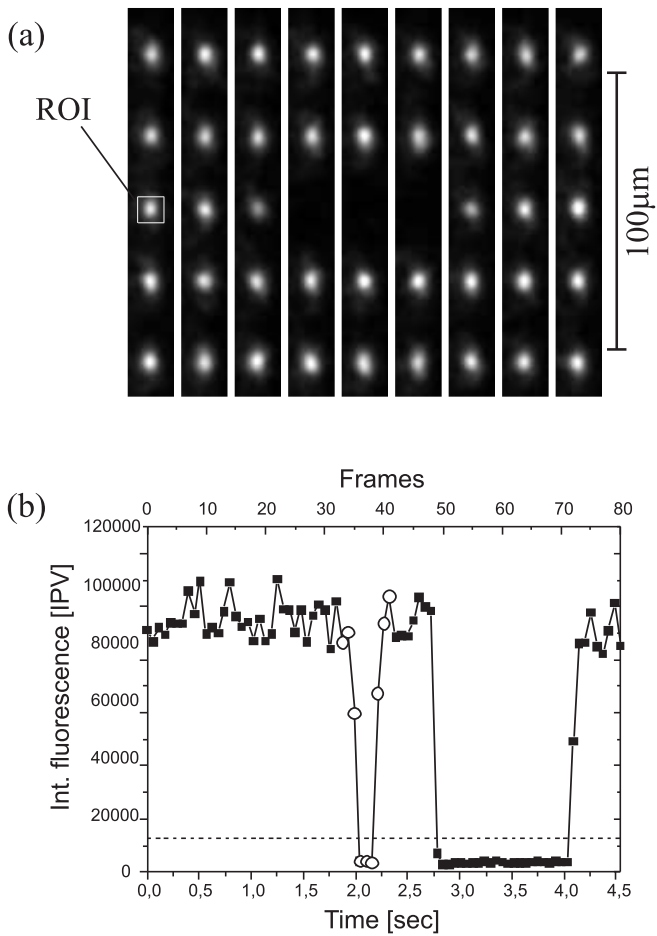}
  \caption{(a) A sequence of CCD-images including a shelving event
  for the central ion. The white square indicates the 'region of interest` (ROI)
 within which the fluorescence is integrated to obtain the signal shown in (b).
 (b) Fluorescence signal from the central ion for an image sequence containing the nine
 images above. The datapoints corresponding to the nine images are indicated by
 open circles. The dashed line indicates the threshold level discussed in the text.}\label{fig:ions_shelving1}
\end{figure}

A sketch of our experimental setup is shown in
Fig.~\ref{fig:setup}. The linear Paul trap used in the experiments
is situated in a stainless steel ultra-high-vacuum chamber. The
trap consists of four cylindrical gold-coated stainless steel rods
arranged in a quadrupole configuration, where two diagonally
opposite rods are separated by 7.00\unit{mm}. The rods are
8.00\unit{mm} in diameter, and each rod is sectioned into three
parts, where the center piece is 5.40\unit{mm} long, and the two
end pieces are 20.00\unit{mm} long. By applying an RF-voltage to
two diagonally opposite electrode rods and applying the same
voltage with opposite phase to the other two electrode rods, we
obtain an effectively harmonic radially confining potential. In
the experiments reported here, the peak-peak amplitude of the
RF-voltage is 600\unit{V} and the frequency is 3.894\unit{MHz},
which results in a radial trap frequency of $\omr\approx 2\pi\cdot
550\unit{kHz}$. Axial confinement is provided by a 580\unit{mV}
DC-voltage on all eight end electrodes, yielding an axial trap
frequency $\omz\approx 2\pi\cdot 50\unit{kHz}$. By adding
additional small DC-voltages to the individual electrodes, we can
carefully center ions in the trap. At the center of the trap a
collimated atomic $Ca$ beam is crossed by a photoionizing laser
beam, and \cafp\ ions are produced by an isotope-selective
resonance-enhanced two-photon ionization
process~\citep{Kjaergaard-ionization, Mortensen-isotope} and
loaded into the trap. The $Ca$ atomic beam originates from a
calcium sample contained in an oven. The $Ca$ atoms effuse out of
a hole in the oven, and by a number of skimmers a collimated
atomic beam is produced. An oven shutter enables us to fully block
the atomic beam. In the present experiments the oven was heated to
$420^{\circ}\unit{C}$, under which conditions the chamber pressure
was measured by an ion gauge~\endnote{A Bayard-Alpert gauge, model
AIG17G from Arun Microelectronics Ltd..} to be $6.0\cdot
10^{-11}\unit{Torr}$. After having loaded five ions and forced
them onto a string by adjusting the trap parameters and applying
Doppler laser-cooling as described below, the oven temperature was
reduced, and the oven shutter was closed. During a one hour
measuring session, the pressure dropped to about $3.6\cdot
10^{-11}\unit{Torr}$. The ions are Doppler laser-cooled on the
$4s\state{2}{S}{1/2}$-$4p\state{2}{P}{1/2}$ transition with
typically 15\unit{mW} of 397\unit{nm} light from a
frequency-doubled Ti:Sapphire laser. From the
$4p\state{2}{P}{1/2}$ state, the ions will not always decay back
to the $4s\state{2}{S}{1/2}$ state but sometimes decay by a
dipole-allowed transition to the $3d\state{2}{D}{3/2}$ state, from
where they are repumped back to the $4p\state{2}{P}{1/2}$ state
using a diode laser at 866nm, see Fig.~\ref{fig:optical-pumping}.
The ions are observed by collecting fluorescence light, emitted at
$397\unit{nm}$ during the Doppler cooling process, with a Nikon
objective lens for MM40/60 measuring microscopes (10x
magnification, f-number$\sim1.7$) placed about 5\unit{cm} above
the trap center (outside the vacuum chamber). The collected light
is amplified by an image intensifier and imaged onto a CCD-camera,
resulting in an all over magnification of 13.5 for the entire
imaging system~\endnote{The image intensifier is from Proxitronic,
model BV 2581 BY-V 1N, and the CCD-camera is a SensiCam system
from PCO.}. The CCD-camera has a 50\unit{ms} exposure time, and
digitial images of the ions with 12-bit resolution are recorded to
the RAM of a personal computer at a frame-rate of 17.610\unit{Hz}.
In order to measure the lifetime of the $3d\state{2}{D}{5/2}$
state, during the Doppler cooling process we excite ions in the
$3d\state{2}{D}{3/2}$ state to the $4p\state{2}{P}{3/2}$ state
using a diode laser at 850\unit{nm}, as shown in
Fig.~\ref{fig:optical-pumping}. From the $4p\state{2}{P}{3/2}$
state the ions can spontaneously decay to the
$3d\state{2}{D}{5/2}$ state. When the ion is in the
$3d\state{2}{D}{5/2}$ state, the optically active electron of the
ion is said to be \textit{shelved}, and the fluorescence on the
397\unit{nm} transition is quenched. The time spent by the ion in
the $3d\state{2}{D}{5/2}$ state before it decays back to the
ground state, will in the following be called a \textit{shelving
period}. By continuously applying the two cooling lasers and the
shelving laser at 850\unit{nm}, we obtain characteristic
fluorescence signals like the one shown in
Fig.~\ref{fig:ions_shelving1}.
\section{Data analysis}\label{sec:Data Analysis}
\subsection{Data reduction}\label{subsec:Data reduction}
The lifetime of the \state{2}{D}{5/2} state can easily be
estimated from the distribution of the shelving periods, since we
expect it to be exponential with a time constant equal to the
lifetime $\tau$.

The duration of a shelving period can be determined by dividing
the number of consecutive frames, where an ion is shelved, by the
frame-rate of the CCD-camera. Our raw data are digital images of
five ions, as shown in Fig.~\ref{fig:ions_shelving1}.a, and our
main dataset consists of $\sim200.000$ such images taken in three
one hour experimental runs. The details of obtaining the
distribution of the shelving periods from these images is
described below.

From the images in Fig.~\ref{fig:ions_shelving1}.a, it is evident
that the ions are spatially well resolved, and that a 'region of
interest` (ROI) around each ion can be defined. The ROI is
13\unit{x}\unit{13} pixels, corresponding to a region of
9.5\unit{\mu m}\unit{x}\unit{9.5}\unit{\mu m} in the trap region.
In order to establish a fluorescence datapoint which reflects the
real ion fluorescence rate within a given image frame, we simply
integrate the pixel values within the ROI. This is a valid
measure, since the image intensifier and the CCD-chip have a
linear response to the fluorescence collected by the objective
lens. In the following the ion fluorescence will be given in
integrated pixel values (IPV).

\begin{figure}
  \centering
  \includegraphics[width=\linewidth]{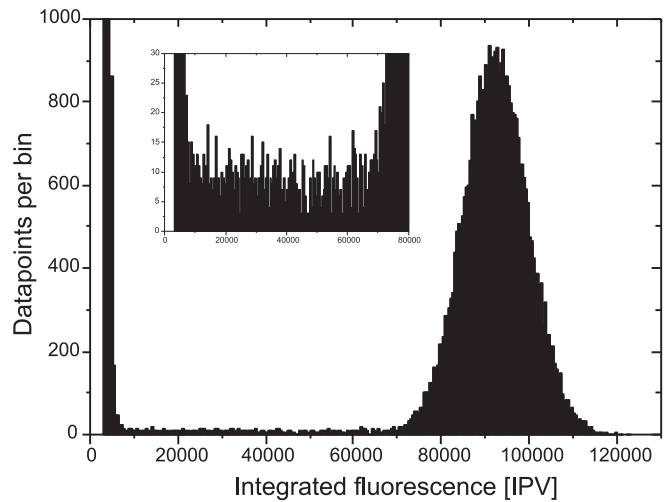}
  \caption{Distribution of datapoints from a single ion, binned
in intervals of 500\unit{IPV}. The maximum of the narrow
background peak near 4000\unit{IPV} is not shown but has a value
of 8000. From the Gaussian distribution around 90000\unit{IPV}
with a width of $\sigma\sim 7500\unit{IPV}$, we can define a
fluorescing level as the center of this distribution. The inset
shows the intermediate points between 0\unit{IPV} and
80000\unit{IPV}.}\label{fig:data_dist}
\end{figure}

As shown in Fig.~\ref{fig:data_dist}, the distribution of the
fluorescence datapoints from a single ion is characterized by a
rather sharp peak at a low fluorescence level and a Gaussian
distribution of datapoints around a level of $\sim
90000\unit{IPV}$, originating from cases where the ion is
scattering 397\unit{nm} light during a whole frame. In the
following we define the fluorescing level as the center of the
Gaussian distribution. The standard deviation of the Gaussian
distribution is $\sigma\thicksim 7500\unit{IPV}$, which is mainly
set by image-intensifier noise, but also by laser intensity and
frequency drifts during the experimental run, and by the finite
ion temperature. The peak at the low fluorescence level, or the
background level, is due to datapoints originating from cases
where the ion is shelved during a whole frame. This background
peak is asymmetric with a sharp left edge, since there is
essentially no 397\unit{nm} light scattered into the ROI's, when
the ions are shelved.

In order to establish a distribution of the shelving periods, we
need to introduce a threshold level to discriminate between
fluorescence datapoints corresponding to frames where the ion is
fluorescing (above the threshold level) or shelved (below the
threshold level). From the distribution of fluorescence datapoints
in Fig.~\ref{fig:data_dist}, it is evident that we can choose a
threshold level which can be used to unambiguously distinguish
between the background level and the fluorescing level of an ion.
The specific choice of threshold level is discussed in the
following.

From the inset of Fig.~\ref{fig:data_dist}, it can be seen that
the distribution of fluorescence datapoints contains some points
with a value of the integrated fluorescence between the background
level and the fluorescing level. These intermediate datapoints
arise from images where the ion is only fluorescing during a
fraction of the exposure time of the CCD-chip. This occurs
naturally when an ion is shelved or decays to the ground state
during the exposure of a single frame. Since a particular choice
of threshold level decides whether an intermediate datapoint is
counted as belonging to a shelving period or not, it affects the
precise distribution of the shelving periods. Fortunately, the
choice of another threshold level on average only adds a constant
amount to all the shelving periods, and therefore the decay rate
extracted from their distribution is not influenced by the choice
of threshold level. This fact allows us to choose a threshold
level in a broad range between the background level and the
fluorescing level.

\begin{figure*}
  \centering
  \includegraphics[width=\linewidth]{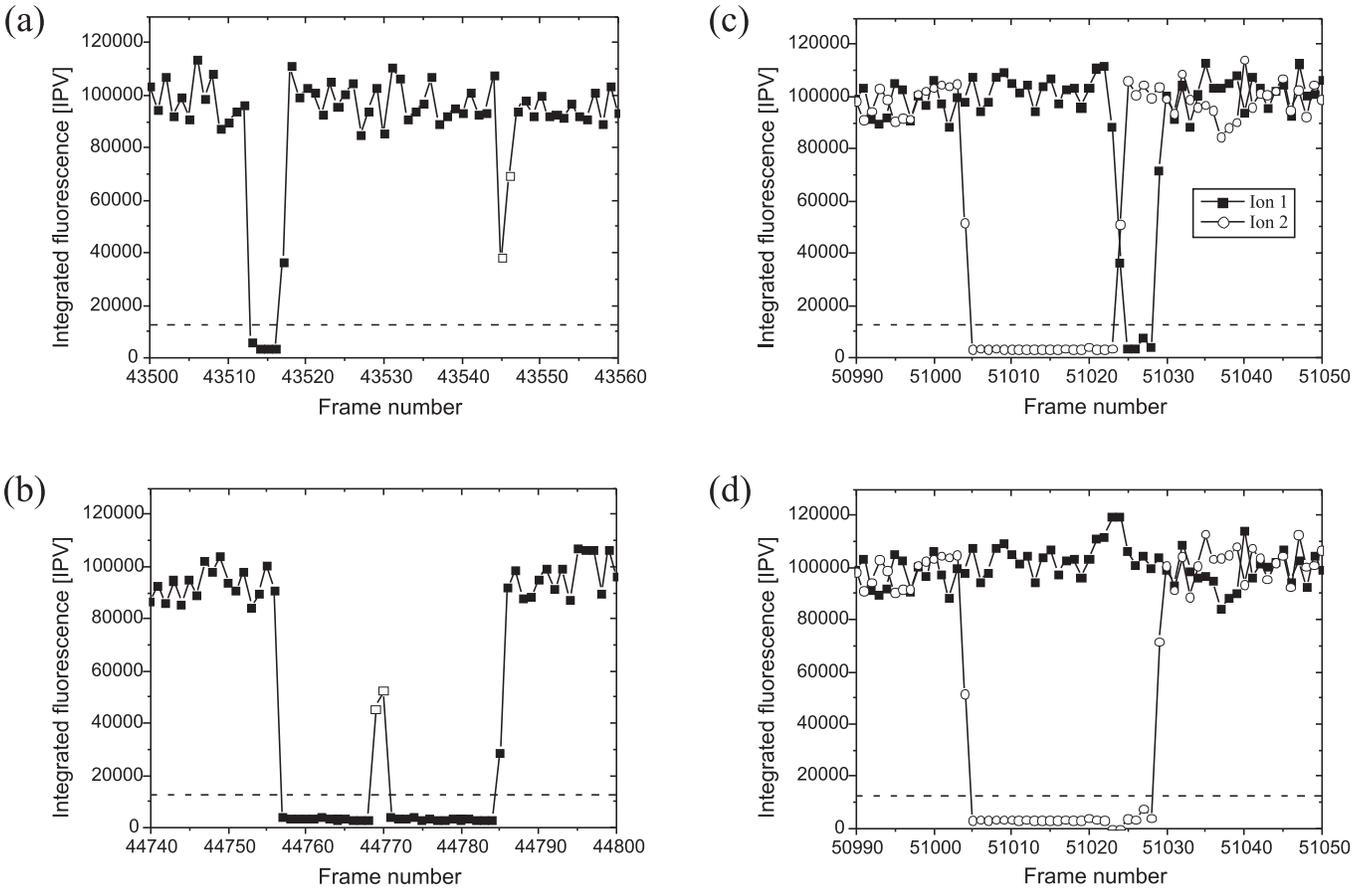}
  \caption{Examples of events which are problematic in connection
 with the data analysis. In all graphs the dashed line indicates the
 chosen threshold level of $\sim 12500$\unit{IPV}. (a) Shelving and fast
 decay; corresponding frames indicated by open squares. (b) Decay and fast
 re-shelving; corresponding  frames indicated by open squares. Here the
 signal does rise above the threshold level, but for an even faster event this may
 not be the case. (c) A shelved and an unshelved ion change places. (d) The
 signals in (c) with the positions of the ions restored, i.e., the signals
 from Ion 1 and Ion 2 are interchanged after the crossing. At the frame where
 the two ions change places and the preceding frame, the fluorescence signal
 from Ion 1 and Ion 2 are given artificial values of 120000\unit{IPV} and 0\unit{IPV}.}
 \label{fig:Reduexamples}
\end{figure*}

Unfortunately, intermediate fluorescence datapoints also occur in
the following three situations: i) shelving followed by fast decay
to the ground state, ii) decay to the ground state followed by
fast re-shelving, iii) a shelved and an unshelved ion change
places, e.g., due to a weak collision with a background gas atom
or molecule or due to the finite temperature of the ions. Examples
of these events, all taking place within one or two camera frames,
are shown in Fig~\ref{fig:Reduexamples}.a-c. Hence, before we
choose the threshold level to be used for extracting the
distribution of shelving periods, we have to consider these three
types of events in some detail.

i) When an ion is shelved at a certain instant of a frame and
decays back to the ground state within that same frame or the
next, then the signal does not necessarily fall \textit{below} the
threshold level, as the example in Fig.~\ref{fig:Reduexamples}.a.
shows. Hence these events may not even be counted as shelving
events. In the data analysis we simply account for such events by
discarding all periods with a duration of one or two frames from
the shelving period distribution. Since the distribution is
expected to be exponential, we can subsequently displace the
remaining distribution, so that periods originally having a
duration of $n$ frames ($n\geq3$) are set to be periods with a
duration of $n-2$ frames.

ii) If an ion decays and is quickly re-shelved, we cannot be
confident that the signal rises \textit{above} the threshold
level, see Fig.~\ref{fig:Reduexamples}.b, and thus two shelving
periods can appear as one longer shelving period, which would
artificially increase the extracted lifetime. Noting that such
events require a \textit{high} shelving rate, it is possible to
exclude them on probability grounds. First, we set a low threshold
level, $T=0.1\cdot(F-B)+B$, where $F$ is the fluorescing level,
and $B$ is the background level, which means that only the very
fastest re-shelving events do not rise above the threshold level.
Second, we require that after a decay, which ends a shelving
period, there must be twenty consecutive datapoints above the
threshold level; otherwise it is not counted as a shelving period
in the data analysis. This requirement is only likely to be
fulfilled with a \textit{low} shelving rate, thus reducing the
risk of accepting shelving periods where a quick re-shelving event
has happened. The probability for an event of decay and quick
reshelving, with the signal not rising above the selected
threshold level, followed by a decay and twenty consecutive
datapoints above the threshold level, is below 2 permille
regardless of the shelving rate. Hence this method gives
\textit{at most} a systematic error of -2\unit{ms} to the final
result, and we will take the systematic error to be $-1\pm
1\unit{ms}$.

iii) As mentioned, two ions may change place, e.g., due to a weak
collision with a background gas atom or molecule.
Fig.~\ref{fig:Reduexamples}.c shows how a shelved and an unshelved
ion changing place, effectively cut one long shelving period into
two shorter ones, which artificially shortens the extracted
lifetime. Therefore, in such events we restore the position of the
ions to obtain a single shelving event, as shown in
Fig.~\ref{fig:Reduexamples}.d. Positions are only restored if the
value of the fluorescence datapoints for the two ions involved
adds up to the fluorescing level $F$ within $\pm 2\sigma$ of this
level. We find about 200 such events in our main dataset. Since
statistically the fluorescence from the two ions does not add up
to the fluorescing level within $\pm 2 \sigma$ in all events, we
estimate a systematic error to the lifetime of $+2\unit{ms}$. True
events of one ion shelving and another decaying within the same
fraction of a frame will, however, occur, and erroneously be
corrected by this procedure. We estimate a systematic error of
$-10\unit{ms}$ due to the erroneously corrected events. All
together the systematic error due to ions switching place is
$-8\unit{ms}$, with an estimated uncertainty of $\pm 4\unit{ms}$.

During data-taking it happened that the ions heated up, so the
ion-string became unstable. Such events are clearly visible on the
images of the ions and were cut out of the dataset before
performing the data reduction process described above. Likewise,
periods where the lasers were adjusted or unstable are not
considered in the further data analysis.

\begin{figure}
  \centering
  \includegraphics[width=\linewidth]{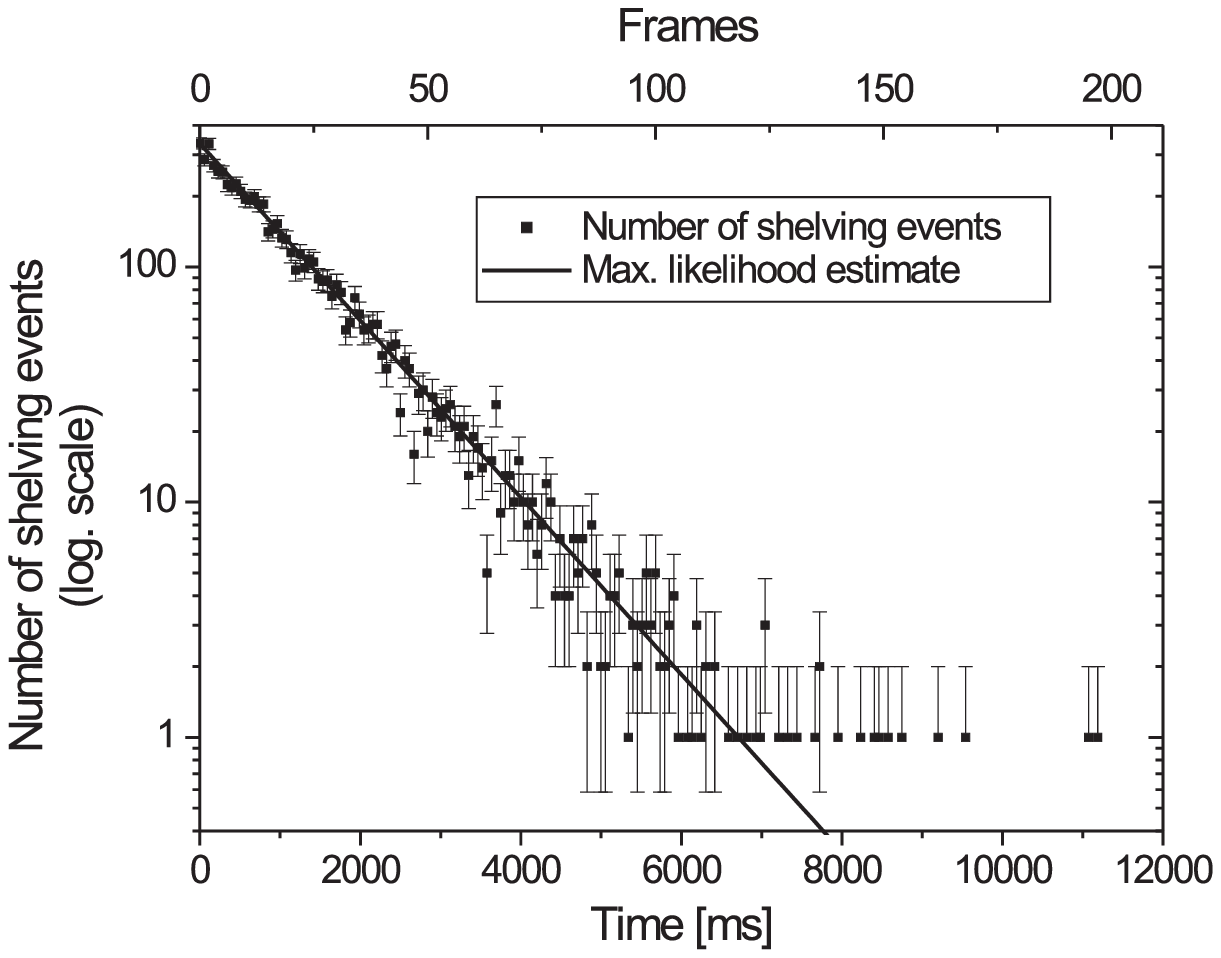}
  \caption{Histogram over the 6805 shelving events obtained after data reduction. Note that the number of shelving events is on a logarithmic scale.
  The error bars are the squareroot of the number of shelving events. The solid line is the maximum likelihood estimate of $\tau=1154\unit{ms}$.}\label{fig:histogram}
\end{figure}

After the data reduction process described above and using a
threshold level of $T=0.1\cdot(F-B)+B$, we extract a distribution
of the shelving periods for each ion, binned in time intervals of
$\Delta t=56.786\unit{ms}$, which is the inverse of the
frame-rate. The counts in equivalent bins for all the ions in the
three experimental runs are then added to obtain a single
distribution, from which a decay rate can be determined. This
distribution is shown in a histogram in Fig.~\ref{fig:histogram},
with the bins shifted such that the histogram has its origin at
time zero.
%
\subsection{Statistical analysis}
From the histogram in Fig.~\ref{fig:histogram}, we can infer the
lifetime $\tau$ of the \state{2}{D}{5/2} state by assuming an
exponential distribution with a decay rate given by the inverse
lifetime of the \state{2}{D}{5/2} state. The histogram is
comprised of 6805 shelving events with duration up to more than 11
seconds. Since, at large times, there are only a small number of
events in the distribution, a least squares fitting method is
inappropriate, and hence we employ a maximum likelihood estimate
instead.

The probability, $p_{i}$, of falling into column $i$ in the
histogram, i.e., having a shelving period of duration $t$, which
fulfills $t_{i}\leq t<t_{i+1}$, where $t_{i}=i\cdot\Delta t$, is
\begin{equation}\label{eq:column-probabibility}
  p_{i}=\int_{t_{i}}^{t_{i+1}}\frac{1}{\tau}e^{-t/\tau}dt=e^{-t_{i}/\tau}\left(1-e^{-\Delta t/\tau}\right).
\end{equation}
Mathematically, the $p_{i}$'s define a proper probability
distribution, since $\sum_{i=0}^{\infty}p_{i}=1$. The sum starts
at $i=0$, corresponding to the origin of the distribution in the
histogram. Since the sum extends to infinity, we must, in
principle, be able to measure infinitely long shelving periods. In
the case of real experiments, where one is only able to measure
periods up to a finite duration, $T_{max}$, the $p_{i}$'s should
formally be renormalized by the factor
$[1-\exp(-T_{max}/\tau)]^{-1}$. In our case where $T_{max}\sim
1\unit{hour}$ and $\tau\sim1\unit{s}$ we can, however, safely
neglect this factor. In order to make the maximum likelihood
estimate, we now introduce the likelihood function
\begin{equation}\label{eq:likelihood}
  L=N!\prod_{i=0}^{\infty}\frac{p_{i}^{n_{i}}}{n_{i}!},
\end{equation}
where $n_{i}$ is the number of shelving events in the $i$'th
column of the histogram, and $N=\sum_{i=0}^{\infty}n_{i}$ is the
total number of shelving events. By maximizing $L$ (or rather $\ln
L$) with respect to $\tau$, we find
\begin{equation}\label{eq:estimate}
  \tau=\frac{\Delta t}{\ln(\frac{\Delta t}{<t>}+1)},
\end{equation}
where ${<t>}=\sum n_{i}t_{i}/N$ is the mean duration of the
shelving periods. The variance of the lifetime
is~\citep{Barlow-statistics}
\begin{eqnarray}\label{eq:variance}
  \rm Var(\tau)=&&-\left(\frac{\partial^{2}\ln L}{\partial\tau^{2}}\bigg\vert_{\frac{\partial\ln L}{\partial\tau}=0}\right)^{-1}=\frac{\tau^{4}(e^{\Delta t/\tau}-1)^{2}}{N(\Delta t)^{2}e^{\Delta t/\tau}}\\\nonumber
  =&&\frac{\tau^{2}}{N}\left(1+\frac{1}{12}\left(\frac{\Delta t}{\tau}\right)^{2}+\mathcal{O}\left[\left(\frac{\Delta
  t}{\tau}\right)^{4}\right]\right).
\end{eqnarray}
Using Eq.~\eqref{eq:estimate} and Eq.~\eqref{eq:variance}, the
lifetime and the statisitical uncertainty can be determined solely
from $\Delta t$, ${<t>}$ and $N$, and we find $\tau=1154\pm
14\unit{ms}$. The exponential distribution based on the maximum
likelihood estimate is plotted as a solid line on top of the
histogram in Fig.~\ref{fig:histogram}.

As a check of the efficiency and validity of our data reduction
process, we have performed a test of goodness-of-fit. To find the
goodness-of-fit we employ the Kolmogorov test, which is based on
the Empirical Distribution Function. A general overview of EDF
statistics can be found in Ref.~\citep{Stephens-GOFtest}, which
also gives tables of significance levels appropriate for cases
like ours, where the lifetime is estimated from the data set. We
find from our data a value for the Kolmogorov test of 0.44,
clearly below the $10\%$ significance level value of 0.995, thus
showing that the fit is good. We note for completeness that our
data set is binned, whereas the values in
Ref.~\citep{Stephens-GOFtest} are for continuous data. However,
since we have more than 100 channels this is not expected to play
any significant role.
\subsection{Radiation, collisions and other potential systematic effects}\label{systematic}
Apart from the systematic errors concerning the data reduction,
already discussed in Section~\ref{subsec:Data reduction}, there
are a few other relevant systematic errors, which will be
discussed in this section.
\subsubsection{Radiation}\label{systematic_radiation}
From the level scheme of Fig.~\ref{fig:optical-pumping}, it is
obvious that any radiation which couples the \state{2}{D}{5/2}
state to the \state{2}{P}{3/2} state may deplete the
\state{2}{D}{5/2} state by excitation to the \state{2}{P}{3/2}
state followed by decay to the \state{2}{S}{1/2} ground state or
the \state{2}{D}{3/2} state and result in a measured lifetime
shorter than the natural lifetime. As a consequence, the
occurrence of such radiation must be considered and, if possible,
reduced. Blackbody radiation at the relevant transition wavelength
is negligible at room temperature, as also discussed by Barton
\textit{et al.}~\citep{Barton}. Other `thermal' sources such as
the ion gauge, roomlight and computer screens were off during the
measuring sessions, except for the screen where we watched the
images of the ions. This screen was facing away from the vacuum
chamber and is hence not expected to cause any problems. Another
class of influencing light sources is the lasers used in the
experiment. In particular the broad background of the emission
spectrum of the 866\unit{nm} and the 850\unit{nm} diode lasers
contains 854\unit{nm} light resonant with
the\state{2}{D}{5/2}-\state{2}{P}{3/2} transition. This radiation
source was first recognized by Block \textit{et
al.}~\citep{Block}, and as noted by Barton \textit{et
al.}~\citep{Barton} the discrepancy between their own and many of
the earlier measurements is possibly due to this previously
unrecognized source of error. To reduce the level of 854\unit{nm}
light from the diode lasers as much as possible, a long-pass
filter was inserted in the 866\unit{nm} beamline and adjusted (by
changing the angle of incidence) to ~30\% transmission at 866nm
and $\lesssim 5\cdot 10^{-4}$ at 854\unit{nm}, and a grating (1200
lines/mm) with a spectrally selective diaphragm (0.9\unit{mm}
diameter, 1080\unit{mm} from the grating) was inserted in the
850\unit{nm} beamline. The filter, the grating and the diaphragm
are shown in Fig.~\ref{fig:setup}. The overall power of the lasers
was reduced to $\sim 18\unit{nW}$ and $\sim 1.7\unit{mW}$ at the
place of the ions, for the shelving laser and the repumping laser,
respectively. The waist size of the beams was $670\unit{\mu
m}\times 700\unit{\mu m}$ for the repumping laser and
$360\unit{\mu m}\times 430\unit{\mu m}$ for the shelving laser.

In order to estimate the lifetime reduction due to radiation
emitted from the diode lasers, we calculate the rate of
de-shelving from the \state{2}{D}{5/2} state, i.e., the excitation
rate from the \state{2}{D}{5/2} state to the \state{2}{P}{3/2}
state times $(1-b)$, where $b=0.068$ is the branching ratio for
decay from the \state{2}{P}{3/2} state back to the
\state{2}{D}{5/2} state~\citep{James}.

First we consider de-shelving due to the repumping laser, which we
split into two contributions, one from the off-resonant
866\unit{nm} radiation and one from the near-resonant background
radiation around 854\unit{nm}. The de-shelving rate due to the
866\unit{nm} radiation was calculated by Barton \textit{et.
al.}~\citep{Barton} to $I_{866}\cdot 9.4\cdot
10^{-5}\unit{s^{-1}/(mW mm^{-2})}$, where $I_{866}$ is the
intensity of 866\unit{nm} radiation. With
$I_{866}=2.3\unit{mWmm^{-2}}$ we find a de-shelving rate of
$2.2\cdot 10^{-4}\unit{s^{-1}}$, yielding a lifetime reduction of
0.3\unit{ms}.

In order to calculate the excitation rate from the
\state{2}{D}{5/2} state to the \state{2}{P}{3/2} state due to
radiation near 854\unit{nm}, we first consider the rate at a given
frequency $\omega_L$, as expressed by Eq.~(2) in
Ref.~\citep{Barton}:
\begin{equation}\label{eq:excitation_rate_Barton}
 R_{12}=\frac{2J_{2}+1}{2J_{1}+1}\frac{\pi^{2}c^{3}}{\hbar\omega_{12}^{3}}A_{21}\frac{I}{c}g(\omega_L-\omega_{12}),
\end{equation}
where $J_{1}=5/2$ and $J_{2}=3/2$ are the total angular momenta of
the involved levels, $\omega_{12}=2\pi
c/854.209\unit{nm}$~\citep{James} is the transition frequency,
$A_{21}=7.7\cdot 10^{6}\unit{s^{-1}}$~\citep{Barton} the Einstein
coefficient for spontaneous decay from the \state{2}{P}{3/2} state
to the \state{2}{D}{5/2} state, $I$ the intensity of the incoming
radiation, and $g(\omega_L-\omega_{12})$ a normalized
Lorentz-distribution. Assuming a flat background spectrum of the
diodelasers, i.e., the intensity per frequency interval $\Delta
I/\Delta\omega_{L}$ is constant, we can integrate
Eq.~\eqref{eq:excitation_rate_Barton} over frequency $\omega_L$
and find an excitation rate of
\begin{equation}\label{eq:excitation rate}
  R=\frac{2J_{2}+1}{2J_{1}+1}\frac{\pi^{2}c^{2}}{\hbar\omega_{12}^{3}}A_{21}\frac{\Delta
  I}{\Delta\omega_{L}}.
\end{equation}

Using a diffraction grating we have measured $\Delta
I/\Delta\omega_{L}\lesssim0.19\unit{nW/(mm^2\cdot GHz)}$ near
854\unit{nm} at the power used in the experiment, and we then find
$R=0.77\unit{s^{-1}}$. Multiplying this rate by $1-b$ and the
transmission of $5\cdot 10^{-4}$ of the long-pass filter, we find
a de-shelving rate of $3.6\cdot10^{-4}\unit{s^{-1}}$, yielding a
lifetime reduction of 0.5\unit{ms}.

As a check of this estimate, we performed an experiment, again
with five ions, with the cooling lasers on but without the
shelving laser. Even without the shelving laser the ions may be
shelved due to radiation from the 866\unit{nm} repumping laser,
which couples the \state{2}{D}{3/2} state to the \state{2}{P}{3/2}
state. From the observed shelving rate, we can then find the
excitation rate on the 850\unit{nm}
\state{2}{D}{3/2}-\state{2}{P}{3/2} transition and compare it to
the calculated excitation rate for the 854\unit{nm}
\state{2}{D}{5/2}-\state{2}{P}{3/2} transition, taking the
different linestrengths into account. In this experiment the power
of the 866\unit{nm} laser was 7.3\unit{mW}, the waist was as
above, the long-pass filter was removed and the oven-shutter was
open. In about 35 minutes we observed 12 shelving events, i.e.,
the observed shelving-rate is $5.8\cdot10^{-3}\unit{s^{-1}}$.
Taking into account the number of ions, the population of the
D-state ($\sim 1/3$, since both cooling transitions are saturated)
and the branching ratio $b$, we find the excitation rate on the
\state{2}{D}{3/2}-\state{2}{P}{3/2} transition: $(3/5b)\cdot
5.8\cdot10^{-3}\unit{s^{-1}}=5.1\cdot 10^{-2}\unit{s^{-1}}$. By
multiplying this number with the relative linestrength between the
\state{2}{D}{5/2}-\state{2}{P}{3/2} transition and the
\state{2}{D}{3/2}-\state{2}{P}{3/2} transition of
8.92~\citep{James}, we find an estimated value of
$0.45\unit{s^{-1}}$ for the excitation rate on the
\state{2}{D}{5/2}-\state{2}{P}{3/2} transition, which should be
compared to $R=0.77\unit{s^{-1}}$ calculated above. The two
numbers are not expected to be equal but only of the same order of
magnitude, since the transition wavelengths are different, and the
diode laser background is not necessarily equally strong near
850\unit{nm} and 854\unit{nm}. Also some of the shelving events
may be due to collisions, as discussed below. Nevertheless, the
numbers agree within a factor of 2 and our calculated estimate of
the excitation rate, and hence the de-shelving rate seems to be
reasonable.

\begin{figure}
  \centering
  \includegraphics[width=\linewidth]{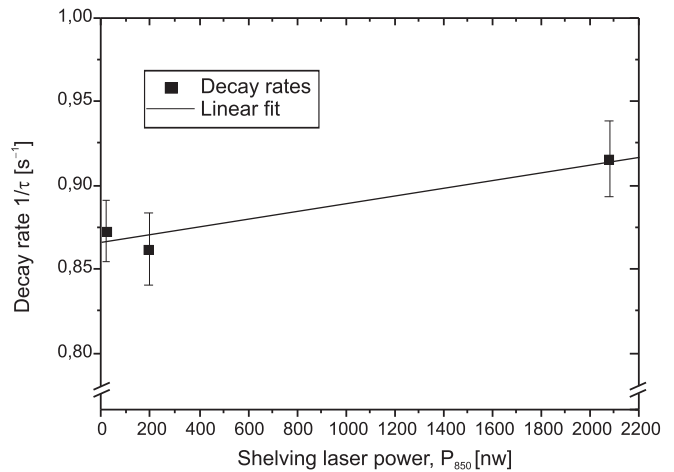}
  \caption{Decay rate measurements at three different power levels of the shelving laser.
   A weighted least-squares linear fit to the measured decay rates yields $1/\tau=0.866(8)\unit{s^{-1}}+0.023(7)\unit{s^{-1}/nW}\cdot P_{850}$.}\label{fig:850power}
\end{figure}

Above we considered de-shelving due to the repumping laser. In
exactly the same way we could consider de-shelving due to the
shelving laser. However, since the intensity of the shelving laser
is much lower than for the repumping laser, and the grating and
the diaphragm strongly reduce the level of 854\unit{nm} light, the
excitation rate is expected to be extremely small. To check this
we performed additional lifetime measurements at three different
power levels of the shelving laser, 18\unit{nW}, 193\unit{nW} and
2081\unit{nW}, yielding lifetimes of $1146\pm24\unit{ms}$,
$1160\pm29\unit{ms}$ and $1092\pm27\unit{ms}$, respectively. When
using higher power, the laser was detuned from resonance to get a
similar shelving rate in all experiments. In
Fig.~\ref{fig:850power}, the linear fit to the decay rates
corresponding to the lifetimes shows that there is a weak
dependence of the decay rate (or lifetime) on the 850\unit{nm}
power. However, with only 18\unit{nW} the lifetime is only reduced
by 0.5\unit{ms}. The results of the two low power measurements and
the crossing at zero power are in agreement with the result
obtained from our main dataset. The measurement at 18\unit{nW} was
in fact performed at the same shelving laser power as the
measurements for the main dataset.

All together, we include a systematic error of $+1\pm 1\unit{ms}$
in our final result due to de-shelving from the two diode lasers.

\subsubsection{Collisions}
Another systematic effect which shortens the measured lifetime is
collisions with background gas atoms and molecules. There are two
relevant types of collisions: fine-structure changing (j-mixing)
collisions and quenching collisions. In a j-mixing collision the
internal state can change from the \state{2}{D}{5/2} state to the
\state{2}{D}{3/2} state, or vice versa. In a quenching collision
the internal state changes from the \state{2}{D}{5/2} state to the
\state{2}{S}{1/2} ground state. Both types of collisions deplete
the \state{2}{D}{5/2} state and hence shorten the measured
lifetime. j-mixing and quenching rate constants ($\Gamma_{j}$ and
$\Gamma_{Q}$) in the presence of different gases are given in
Ref.~\citep{Knoop-collisions} and references therein. Quite
generally j-mixing collisions are found to be an order of
magnitude stronger than quenching collisions. From a restgas
analysis~\endnote{Spectra restgas analyzer with LM61 satellite,
LM502 analyzer and LM9 RF-head.}, we know that the restgas in our
vacuum chamber is mainly composed of $\rm H_{2}$ and gas of 28
atomic mass units, i.e., $\rm N_{2}$ or $\rm CO$. In the restgas
analysis it was not possible to distinguish between $\rm N_{2}$
and $\rm CO$, since the chamber pressure was so low that the
signal from the atomic constituents of these molecules could not
be observed. Knoop \textit{et al.}~\citep{Knoop-collisions} found
the following rate constants in units of \unit{cm^{3}s^{-1}} for
collisions with $\rm H_{2}$ and $\rm N_{2}$: $\Gamma_{j}(\rm
H_{2})=(3\pm 2.2)\cdot 10^{-10}$, $\Gamma_{Q}(\rm H_{2})=(37\pm
14)\cdot 10^{-12}$, $\Gamma_{j}(\rm N_{2})=(12.6\pm 10)\cdot
10^{-10}$ and $\Gamma_{Q}(\rm N_{2})=(170\pm 20)\cdot 10^{-12}$.
The measurements were performed on a cloud of relatively hot ions,
as compared to the laser cooled ions considered in this paper. As
noted in Ref.~\citep{Knoop-collisions}, other measurements of
j-mixing with $\rm H_{2}$ at different collision energies give
similar results, so we may expect only a weak energy-dependence
for the j-mixing collisions, and therefore we use the values given
in Ref.~\citep{Knoop-collisions}. On the other hand, higher
quenching rates are expected at lower
temperatures~\citep{Knoop-collisions}, but to our knowledge there
are no measurements of that for $\rm Ca^{+}$. Unfortunately, we
are not aware of any similar measurements with CO, and in our
estimate of the collision rate below, we therefore assume that the
mass 28 restgas is $\rm N_2$. The relatively large rate constants
of $\rm N_{2}$ found in Ref.~\citep{Knoop-collisions} indicate
that at least this assumption probably does not lead to a large
underestimate of the collision induced de-shelving rate.

Assuming that the values given in Ref.~\citep{Knoop-collisions}
are applicable, we can estimate the collision-induced lifetime
reduction. At a pressure of $5\cdot 10^{-11}\unit{Torr}$, taking
into account the sensitivity to different gases of the ion gauge
and using the result of the restgas analysis, we find that the
restgas is composed of $54\%\rm H_{2}$ and $46\%\rm N_{2}$, which
yields a total collision induced de-shelving rate of $2.3\cdot
10^{-3}\unit{s^{-1}}$, and a systematic error to the lifetime of
+3\unit{ms} with an estimated uncertainty of $\pm 1\unit{ms}$ .

From the measurement of 12 shelving events in 35 minutes without
the shelving laser on, as described in
Sec.~\ref{systematic_radiation} above, we can obtain an
\textit{upper limit} for the j-mixing collision rate if we assume
that all the observed shelving events are due to j-mixing
collisions inducing a transition from the \state{2}{D}{3/2} state
to the \state{2}{D}{5/2} state (with rate $\gamma_{35}$). Again
taking the number of ions and the population of the
\state{2}{D}{3/2} state into account, the collision induced
shelving rate is $3.5\cdot 10^{-3}\unit{s^{-1}}$. Since the oven
shutter was open in that experiment, thus allowing collisions with
neutral calcium atoms as well, we expect this number to be larger
than under the conditions for the `real' lifetime measurements.
The transition rate, $\gamma_{53}$, for the
\state{2}{D}{5/2}-\state{2}{D}{3/2} de-shelving transition is
expected to be given by $\gamma_{53}=2\gamma_{35}/3$, owing to the
principle of detailed balance, so the collision induced
de-shelving rate is $2.3\cdot 10^{-3}\unit{s^{-1}}$. This is our
upper limit for the j-mixing collision rate, which should be
compared to our estimate above of the \textit{total}
collision-induced de-shelving rate, which is dominated by
contributions from j-mixing collisions. Somewhat fortuitously, the
numbers are equal, and hence our calculated estimate seems to be
reasonable.
\subsubsection{Other effects}
When observing shelving events from a string of ions, one might
consider if there are any correlations in the decay of the
individual ions from the \state{2}{D}{5/2} state to the ground
state, which could influence the measured lifetime. In the
experiments by Block \textit{et al.}~\citep{Block} indications of
correlated decays from the \state{2}{D}{5/2} state were observed,
manifested as an overrepresentation of events where several ions
decay at the same time. A later detailed experiment by Donald
\textit{et al.}~\citep{Donald-correlations} showed, however, no
such correlations. Apart from sudden bursts of 854\unit{nm}
radiation, the only reasonable physical mechanism which could lead
to correlations is so-called subradiant and superradiant
spontaneous emission due to interference in the spontaneous decay
of two or more ions~\citep{DeVoe-superradiance}. In the simple
case of two ions, superradiant and subradiant spontaneous emission
is characterized by a relative change in the normal (single ion)
decay rate of the order of $\pm\sin(kR)/kR$ (when $kR>10$), where
R is the ion-ion distance, $k=2\pi/\lambda$ with
$\lambda=729\unit{nm}$ in our case~\endnote{In
Ref.~\citep{DeVoe-superradiance} a dipole transition is
considered. For an electric quadrupole transition the relative
change of the decay rate is of the same order of magnitude.}. For
more ions the effect is of the same order of magnitude. In our
case $1/kR\approx6\cdot 10^{-3}$ so just from this argument the
effect is small, but not negligible. The interference effect,
however, relies on creating and maintaining a superposition state
of the form
$\ket{\pm}=(\ket{S_{1}D_{2}}\pm\ket{D_{1}S_{2}})/\sqrt{2}$, where
$S$ and $D$ indicate the internal state, \state{2}{S}{1/2} or
\state{2}{D}{5/2}, and indices 1 and 2 relate to Ion 1 and Ion 2.
In our experiment such a superposition state can only be created
by a random process since the \state{2}{D}{5/2} state is only
populated through spontaneous emission, and consequently the
relative phase between the states \ket{S_{1}D_{2}} and
\ket{D_{1}S_{2}} is expected to be random. This fact would in
itself average out the effect on the lifetime. In addition, if the
superposition state is created, it is immediately (as compared to
$\tau$) destroyed since, from a quantum mechanical point of view,
we are constantly measuring the internal state of the ions. So any
super- or subradiant effect is in fact expected to be destroyed,
and we do not expect any correlation in decays from the
\state{2}{D}{5/2} state. We have checked our data for correlated
decays, and indeed a statistical analysis shows no evidence of
correlations.

In Ref.~\citep{Barton} mixing of the \state{2}{D}{5/2} state with
the \state{2}{P}{3/2} state due to static electric fields was
considered and found to be negligible. For our trap we also find
this effect to be negligible.

Finally, we note that the read-out time of the camera influences
the measured duration of the shelving periods. As for the choice
of threshold level, the read-out time has no effect on the
measured decay rate and hence on the measured lifetime.
\subsubsection{The total effect of systematic errors}
Above we have identified and evaluated systematic errors
originating from the data analysis and from de-shelving due to
radiation and collisions. An overview of the estimated errors and
their uncertainties is given in Table~\ref{Table:sys_errors}.

The effects leading to the systematic errors can be modelled by
extra decay rates added to the natural decay rate:
\begin{align}\label{eq:SysErrors}
  &\frac{1}{\tau_{meas}}=\frac{1}{\tau_{nat}}+\sum_{i}\gamma_{i}
  \intertext{or}
  \label{eq:SysErrors2}
  &\tau_{nat}\approx\tau_{meas}+\sum_{i}\tau_{meas}^{2}\gamma_{i},
\end{align}
where $\tau_{meas}$ is the measured lifetime as determined from
the maximum likelihood estimate, $\tau_{nat}$ is the natural
lifetime, and the $\gamma_{i}$'s are the extra decay rates, which
can attain both positive and negative values in this model. The
systematic errors given in the text and Table I correspond to
$\tau_{meas}^{2}\gamma_{i}$. Eq.~\eqref{eq:SysErrors2} shows that
the systematic errors should be added linearly, yielding
-5\unit{ms}, and added to the result of the maximum likelihood
estimate, giving a lifetime of 1149\unit{ms}. The uncertainties of
the systematic errors are independent, and therefore they are
added quadratically, yielding an uncertainty of $\pm 4$\unit{ms}.
\begin{table}
\centering
\begin{tabular}{lc} \hline\hline
 Effect&  Systematic error [\unit{ms}]\\\hline
  Quick re-shelving & $-1\pm1$  \\
  Ions switching place & $-8\pm 4$  \\
  De-shelving, diode lasers & $+1\pm 1$  \\
  De-shelving, collisions & $+3\pm 1$  \\\hline
  Total & $-5\pm4$ \\\hline\hline
\end{tabular}
\caption{Overview of estimated systematic
errors.}\label{Table:sys_errors}
\end{table}
\section{Result and conclusion}\label{sec:conclusion}
By correcting the maximum likelihood estimate with -5\unit{ms}, as
described above, we find that our final result for the lifetime
measurement is
\begin{equation}\label{eq:tau}
  \tau_{nat}=1149\pm{14}(\rm stat.)\pm 4(\rm sys.)\unit{ms}.
\end{equation}
The largest error is the statistical, but we do have a
non-negligible systematic uncertainty, originating from the
correction procedure when ions change places. The associated
error, and hence the uncertainty, could be reduced by increasing
our signal-to-noise ratio, which would narrow the time window
where real events of simultaneous decay and shelving for two
different ions could be taken for two ions changing place. The
signal-to-noise ratio can be improved by frequency locking the
Ti:Sapphire laser and power stabilizing the output from the
doubling cavity. From Eq.~\eqref{eq:variance} we see that the
statistical uncertainty can only be reduced by a longer data
acquisition time (increasing N), and not simply by increasing the
frame-rate, since the second term in the expansion is already
negligible in our case.

Our measurement was performed with a string of ions, unlike most
other shelving experiments. In all single ion experiments, one has
to consider the fact that the ion may heat up, so the fluorescence
level drops significantly, and the ion can appear to be shelved,
although it is not. In our experiment with five ions on a string,
we can detect and discard all events of this kind since if one ion
or several ions heat up, we would see the remaining ions move or
heat up as well. Moreover, with a string of ions, the shelved ions
are sympathetically cooled by the unshelved ions, so the number of
heating events are expected to be reduced using a string of ions,
as compared to single ion experiments. The major drawback of using
a string of ions is that ions may change places, which influences
the measured lifetime if this is not taken care of in the data
analysis, as demonstrated here.

In conclusion, we have measured the lifetime of the metastable
3d\state{2}{D}{5/2} state in the \cafp\ ion using the shelving
technique on a string of five ions. Our result agrees roughly at
the level of one standard deviation with the already mentioned
result obtained by Barton \textit{et al.}~\citep{Barton}, with the
value reported in Ref.~\citep{Knoop-lifetimeII} and with two
theoretical papers~\citep{Vaeck-lifetime, Brage-lifetime}. On the
level of two standard deviations, the result agrees with two other
measurements where de-shelving due to diode lasers was taken into
account~\citep{Donald-correlations, Block} and with the
storage-ring measurement by Lidberg \textit{et
al.}~\citep{Lidberg-lifetime}. The most recent measured
values~\citep{Barton,Knoop-lifetimeII}, including our result, are
in good agreement and group themselves around a value of about
1160\unit{ms} within 1\% of that value. This newly obtained level
of agreement should provide valuable input to future atomic
structure calculations and astrononomical studies.

\section{Acknowledgements}
The authors gratefully acknowledge Karsten Riisager for valuable
discussions about the statistical analysis of the data. This work
was supported by QUANTOP - the Danish National Research Foundation
Center for Quantum Optics and the Carlsberg Foundation.
\bibliography{staanum}
\end{document}